\documentclass[prl,twocolumn,aps,showpacs]{revtex4}
\usepackage{graphicx}
\usepackage{amsmath}
\usepackage{subfigure}
\begin{document}

\title{Tunneling in suspended carbon nanotubes assisted by longitudinal phonons}

\author{S. Sapmaz, P. Jarillo-Herrero, Ya. M. Blanter, C. Dekker, and H.S.J. van der Zant}
\affiliation{ Kavli Institute of Nanoscience, Delft University of
Technology, PO Box 5046, 2600 GA Delft, The Netherlands}
\date{\today}
\begin{abstract}
Current-voltage characteristics of suspended single-wall carbon
nanotube quantum dots show a series of steps equally spaced in
voltage. The energy scale of this harmonic, low-energy excitation
spectrum is consistent with that of the longitudinal low-$k$
phonon mode (stretching mode) in the nanotube. Agreement is found
with a Franck-Condon-based model in which the phonon-assisted
tunneling process is modeled as a coupling of electronic levels to
underdamped quantum harmonic oscillators. Comparison with this
model indicates a rather strong electron-phonon coupling factor of
order unity.
\end{abstract}

\maketitle

In nano-electromechanical systems (NEMS), mechanical motion
affects electrical current and vice
versa~\cite{Roukes,Cleland,Shekhter,Blick}. Of special interest is
the study of electron-phonon coupling in these devices since
tunneling of a single electron may induce a displacement of the
movable
structure~\cite{Flensberg,Tuominen,Blanter,Belzig,Mitra,von
Oppen}. The interaction strength is characterized by the
dimensionless electron-phonon (e-ph) coupling constant $g$, which
is proportional to the ratio of the classical and the quantum
displacement. In bulk systems the e-ph coupling is generally weak
and the coupling constant is orders of magnitude smaller than one.
However, since the coupling dramatically increases with decreasing
device mass, NEM-devices may exhibit an intermediate to strong
e-ph coupling~\cite{Park C60,C140,Kotthaus}. In this regime,
current-voltage characteristics are expected to exhibit additional
steps whose height can be used as an estimate of $g$. For example,
$g$ is around one in the C$_{60}$ molecular devices of
Ref.~\cite{Park C60}, while measurements on different C$_{140}$
samples~\cite{C140} indicate a value of $g$ between 0.2 and 8.

Carbon nanotubes (NTs) are ideal systems for exploring
electro-mechanical effects since they have a small diameter, a low
mass, and can be defect free on a molecular level. In experiments
on suspended nanotubes, different methods have already been used
to probe the bending~\cite{Reulet,Sazanova} and radial breathing
mode (RBM)~\cite{RBM Brian}. The measurements show that the
free-hanging tubes operate in the underdamped regime of low
dissipation. For the fundamental bending mode the reported quality
factor is about 100; for the RBM it is estimated as high as 10000.

In this Letter we present electronic transport spectroscopy
measurements on suspended single-wall nanotubes, which show
signatures of phonon-assisted tunneling, evidenced by the presence
of a series of steps in the $I-V$ characteristics. Such steps form
a harmonic low-energy spectrum, whose energy scale and length
dependence are consistent with that of the longitudinal stretching
mode. Comparison with the Franck-Condon theory shows that the e-ph
coupling constant is of order one.

\begin{figure}[ht]
\includegraphics[angle=0,width=0.7\linewidth]{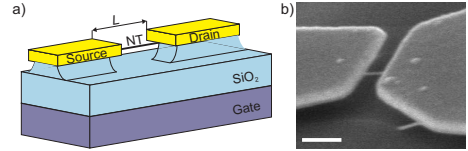}
\caption{\label{fig1} a) Schematic drawing of a suspended nanotube
clamped between two Cr/Au electrodes on top of silicon oxide. The
underlying oxide is partially removed by a wet etch step leaving
the nanotube suspended. The highly doped silicon plane is used as
a global gate to tune the electrostatic potential of the nanotube.
b) Scanning-electron microscope micrograph of a suspended
nanotube. The scale-bar represents 200~nm.}
\end{figure}

Devices are fabricated by locating individual nanotubes (laser
ablation and CVD) on a Si/SiO$_2$ substrate using an atomic force
microscope (AFM) with respect to predefined markers. Subsequently,
the electrodes are made using conventional e-beam lithography
techniques and thermal evaporation of Cr (5~nm) and Au (50~nm).
The nanotubes are suspended by removing the underlying SiO$_2$ in
a wet etch step using buffered HF~\cite{Nygard}. A schematic
sample geometry and SEM micrograph are shown in Fig.~1. In the
experiment the source and gate voltage are defined with respect to
the drain, which is connected to ground.

In Fig.~2 we show stability diagrams for three nanotubes measured
at 10~mK (a) and 300~mK (b,c) where the differential conductance,
$dI/dV$, is plotted versus bias and gate voltage. The three
metallic nanotubes have a length between source and drain
contacts, $L$, ranging from 0.14 to 1.2~$\mu$m. Their diameter $d$
is between 1 and 1.4~nm as determined from AFM imaging. In the
diamond shaped regions (Coulomb diamonds) the current is zero due
to Coulomb blockade, and the charge number in the dot is fixed.
Regular and closing Coulomb diamonds indicate single dot behavior
\cite{Grabert,Leo_Rep} in all three samples for the gate range
shown. Notice that the diamonds in Fig.~2a do close, as shown in
the inset, which was taken at a higher temperature (300~mK) in a
different cooldown. The low-bias current, however, is suppressed
which could be a signature of strong electron-phonon
coupling~\cite{Flensberg,Mitra,von Oppen}.

\begin{figure}[t]
\includegraphics[angle=0,width=0.8\linewidth]{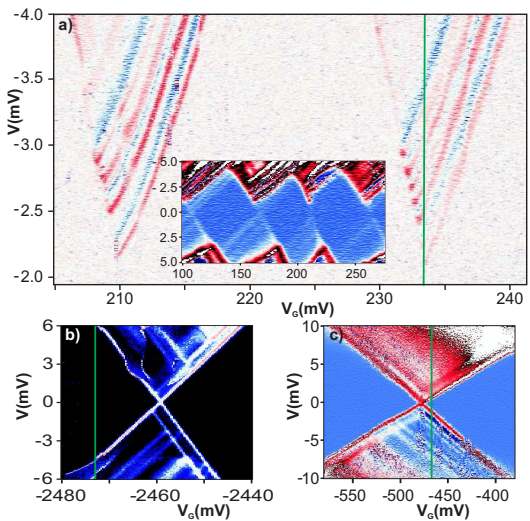}
\caption{\label{fig2} Stability diagrams for three different
suspended nanotubes with a length in between contacts of
1.2~$\mu$m, 420~nm, and 140~nm for (a), (b), and (c) respectively.
The conductance ($dI/dV$) is plotted as a function of source-drain
voltage, $V$, and gate voltage, $V_G$. Blue corresponds to low and
red to high conductance. Measurements have been performed at
$T$=300~mK except in (a), where the base temperature was 10~mK.
(a) Small region of a stability diagram showing closely spaced
sets of lines running parallel to the Coulomb diamond edges for
two charge states. At low bias, a strong suppression of the
conductance is present. Red lines indicate positive differential
conductance; blue lines negative differential conductance. Inset:
regular diamonds that close are observed in a different cool down
at $T$= 300~mK. (b) and (c) Diamond crossings for two other
samples, again showing lines parallel to the diamond edges with
energy separations smaller than expected for electronic
excitations.}
\end{figure}

Excitations of a quantum dot appear as lines running parallel to
the Coulomb diamond edges in the stability
diagrams~\cite{Grabert}. At such a line, a new electronic level
becomes resonant with the leads and an additional transport
channel opens up. The energy of an excitation can be determined by
reading off the intersection point between the excitation line and
the Coulomb diamond edge on the bias axis~\cite{Leo_Rep}.
Furthermore, the excitations correspond to the charge state of the
Coulomb diamond they end up in. Electronic excitations in
nanotubes typically differ between adjacent charge
states~\cite{Cobden}. In Fig.~2a, a dense set of equally spaced
excitation lines (starting from the first electronic excitation)
is clearly visible near $V_G=210$~mV and 230~mV, i.e., adjacent
charge states exhibit a similar set of excitations with
approximately the same energy spacing. The fact that excitations
occur primarily in one direction is due to asymmetric tunnel
barriers~\cite{Leo_Rep}.

The energy differences between the excitation lines of Fig.~2 are
shown in the insets of Fig. 3. In all three cases, the excitation
energy is an integer multiple of the first (fundamental)
excitation. Thus, they form a harmonic spectrum with up to 5
levels. A linear fit yields an excitation energy of 140, 690, and
530~$\mu$eV for the tubes with length $1.2$~$\mu$m (a), 420~nm
(b), and 140~nm (c), respectively. These values are an order of
magnitude smaller than the expected mean electronic level spacing
given by $\Delta=hv_F/2L$ with $h$ the Planck constant and
$v_F=8.1 \cdot 10^5$~m/s~\cite{Serge} the Fermi velocity.

\begin{figure*}[t]
\includegraphics[angle=0,width=0.8\textwidth]{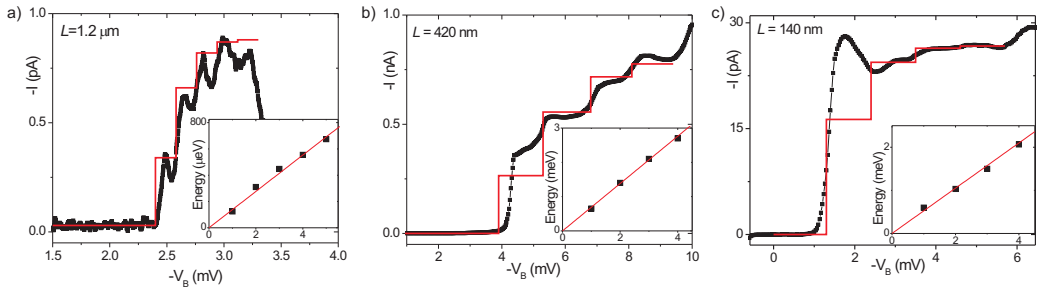}
\caption{\label{fig3} Current as a function of source-drain
voltage at a voltage indicated by the green lines in Fig.~2. The
red lines represent the step heights calculated in the
Franck-Condon model (see text) for an electron-phonon coupling
parameter of 0.95, 1.1, and 0.5 for (a), (b), and (c)
respectively. In the insets, the energy separation between the
peaks or steps (lines in Fig. 2) is plotted, showing equally
spaced, harmonic spectrum. The slope of the drawn line is 140,
690, and 530~$\mu$eV for the insets of (a), (b), and (c)
respectively.}
\end{figure*}

A more natural explanation for the observed harmonic spectra is a
vibrational mode coupled to electron tunneling~\cite{Park C60}.
Multiple steps with identical spacing would than arise from the
excitation of an integer number of vibrational quanta. Indeed, the
observed equidistant energy separation is consistent with that
expected from the longitudinal stretching mode in the nanotubes.
In Fig.~4, we plot the energy of important low-energy vibrational
modes of single-wall nanotubes~\cite{DresselhausAndoEgger,Note}.
For comparison, we plot the mean electronic energy level
separation, $\Delta$, in black. Blue squares correspond to the
fundamental vibrational excitation energy extracted from the
linear fits in the insets of Fig.~3. The energy of the radial
breathing mode (green) does not depend on the nanotube length and
equals 28~meV/$d$(nm). The bending mode (red) has a $L^{-2}$
dependence~\cite{Sapmaz}, and an energy much smaller than the
measured excitation energy. The stretching mode vibration energy
(blue) is inversely proportional to the
length~\cite{Landau-Lifshitz}, $E=(nh/L) \sqrt{Y/\rho_m}$, where
$Y$ is Young's modulus, $\rho_m$ is the density and $n$ is the
vibrational quantum number. For nanotubes with
$\rho_m=1.3$~g/cm$^3$, $Y=1$~TPa the vibrational energy
corresponding to the fundamental mode is $\sim 110~\mu
\text{eV}/L(\mu \text{m})$ \cite{Note}. As Fig.~4 shows, the data
are in good agreement with these predicted values.

\begin{figure}[ht]
\includegraphics[angle=0,width=0.7\linewidth]{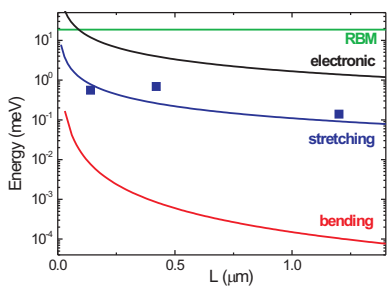}
\caption{\label{fig4} Energy scales of different vibrations and
electronic excitations plotted on a log scale for a nanotube with
a 1.4~nm diameter. The radial breathing mode (green) does not
depend on the length $L$. The bending mode vibrations(red) have a
$L^{-2}$ dependence. The mean electronic level spacing (black) and
the stretching mode (blue) vibrations depend inversely on the
length.}
\end{figure}

The coupling of electronic levels with vibrational modes (quantum
harmonic oscillators) can be described in terms of the
Franck-Condon model~\cite{Chemistry book}. According to the
Franck-Condon principle, an electron in an electronic transition
moves so fast that the nuclear positions are virtually the same
immediately before and after the transition. As a consequence, the
transition rate is proportional to the Franck-Condon factors
defined as the square of the overlap integral between the
vibrational wavefunctions of the two states involved. An important
parameter is the electron-phonon coupling factor,
$g=\frac{1}{2}(\frac{x}{x_0})^2$. This is the ratio of the
classical displacement length, $x$, to the quantum mechanical
oscillator length, $x_0=\sqrt{\hbar/m\omega}$. Alternatively,
$g=\frac{F^2}{2 \hbar m \omega^3}$, where $F$ is the force on, $m$
the mass of, and $\omega$ the frequency of the oscillator.

For low damping, the vibrational levels remain sharp and the
Franck-Condon model predicts steps in the current-voltage
characteristics, that are equally spaced in energy (bias voltage).
In the presence of strong relaxation, the normalized step heights
are given by~\cite{Flensberg}: $P_n=e^{-g}g^n/n!$. In the strong
coupling ($g \gg 1$) limit, the height of the first steps is
exponentially suppressed (phonon
blockade)~\cite{Flensberg,Mitra,von Oppen}. Multiple steps only
arise if $g$ is of the order of one or larger and the observation
of a spectrum of equally spaced excitation lines therefore
indicates that the e-ph coupling in our suspended nanotubes must
be rather strong.

In Fig.~3, the red curves represent the step heights ($P_n$) given
by the Franck-Condon model with strong relaxation discussed above.
The symbols are the experimental curves taken at the green lines
in Fig.~2. Considering the simplicity of the model, reasonable
agreement is obtained in all three cases. The comparison yields an
estimate of $g$ of 0.95, 1.1, and 0.5 in Fig.~3a, b, and c
respectively, indicating that it is approximately length
independent. We have also performed a similar analysis at other
gate voltages yielding the same $g$-values.

The theoretical curves in Fig.~3 do not exactly follow the
measured ones. Better fits may be obtained if the influence of a
gate voltage and asymmetric coupling is
considered~\cite{Flensberg} or if coupling to excited electronic
states~\cite{Maarten} is considered or if the influence of damping
or non-equilibrium phonons (weak relaxation) is taken into
account. In the latter case the peak heights are expected to
display a non-systematic dependence on $g$ and peak
number~\cite{Mitra}. Consideration of these effects is, however,
beyond the scope of this paper.

The high value of the e-ph coupling ($g \sim 1$) is remarkably
since in graphite the coupling between electrons and longitudinal
phonons is weak. A source of intermediate to strong coupling could
be the interaction between longitudinal and transverse
vibrations~\cite{note-flensberg}. Alternatively, we find that in
suspended nanotubes the same e-ph coupling mechanism as in the
bulk~\cite{Abrikosov} can lead to a $g \sim 1$ if the electron
density is inhomogeneous. The calculation proceeds as follows: The
interaction energy of electrons with the polarization charge is
characterized by the energy
\begin{equation}\label{energy}
W=-\int dx dx' \rho(x) K(x-x') \frac{\partial P}{\partial x'}.
\end{equation}
Here, $\rho(x)$ is the density of excess charge produced by one
electron, $K(x - x')$ is an interaction kernel, which we
approximate by $\delta(x - x')$ for the case that interactions are
effectively screened by the gate, and $P(x) \approx e\rho_0 z(x)$
is the polarization vector. The quantity $\rho_0 \sim k_F$ is the
total electron density and $z(x)$ is the displacement, which in
the single-mode approximation becomes $z(x) = A_n \sin (\pi n
x/L)$. Calculating the force $F = -\partial W/\partial A_n$, we
obtain
\begin{equation}\label{ephon}
F = \frac{e\rho_0 \pi n}{L} \int_0^L dx \ \rho(x) \cos \frac{\pi n
x}{L} .
\end{equation}
If the excess charge density is uniform, $\rho(x) = e/L$, $F=0$
for all modes. Incorporation of interactions that are
screened at distances longer than the distance to the gate, yields
a force that scales as $L^{-2}$. In this case, the coupling
parameter $g$ also scales as $L^{-2}$ and typical values are in
the order of $10^{-3}$, in apparent contradiction with the
experimental data.

Assuming that the charge is localized in the center of the tube,
$\rho(x) = e\delta(x-L/2)$, the force is zero for odd harmonics,
but for even harmonics, $n = 2l$, it reads $F_l = (-1)^l e^2
\rho_0 2\pi l/L$. This results in a coupling parameter that is
length independent and scales as $l^{-1}$: higher modes are
coupled weaker to electrons. Numerical estimates show that $g \sim
1$. Localization of an electron in a point away from the middle
produces coupling to both odd and even modes. Note that the
electron does not have to be strongly localized to produce a $g
\sim 1$. Such a non-uniform density can be created by impurities
located in the substrate, or induced by a redistribution of
electrons in a suspended tube bent by a underlying gate
electrode~\cite{Fedorets}.

An interesting feature of the data is the appearance of negative
differential conductance (NDC) in the current-voltage
characteristics. NDC is very pronounced in Fig.~3a, but also
present in Fig.~3b,c. Although several explanations for NDC have
been put forward, its origin remains unclear. Koch and von
Oppen~\cite{von Oppen} showed that for low relaxation and strong
e-ph coupling, NDC features appear, although they do not follow
regions with strong positive differential conductance (PDC) as in
our data. McCarthy {\it et al.}~\cite{Tuominen} have shown that
NDC features can be due to an e-ph coupling that is voltage
dependent. Their calculations also reproduce the catastrophic
current decrease of Fig.~3a for bias voltages higher than 3~mV.
However, at the moment we do not know how such a voltage
dependence would arise in suspended tubes. Finally, Nowack and
Wegewijs~\cite{Maarten} have considered a Franck-Condon model with
a coupling to an electronic and its excited state. They show that
the competition between the two states generates strong NDC
effects. NDC and PDC lines may have the same gate voltage
dependence preceded by a region of suppressed current. This
scenario may especially be relevant for the data in Fig. 3a.

In summary, transport measurements on suspended SWNTs show
signatures of phonon-assisted tunneling, mediated by longitudinal
vibrational (stretching) modes. The current-voltage
characteristics show multiple steps whose heights are in
reasonable agreement with the Franck-Condon predictions if the
e-ph coupling constant is of order unity. Suspended nanotube
quantum dots form an interesting model system for future studies
on the interaction between single electrons and quantized phonons
in the intermediate to strong electron-phonon coupling limit.

We thank Karsten Flensberg, Wataru Izumida, Leo Kouwenhoven,
Leonid Gurevich and Maarten Wegewijs for discussions. Financial
support is obtained from the Dutch organization for Fundamental
Research on Matter (FOM), which is financially supported by the
'Nederlandse Organisatie voor Wetenschappelijk Onderzoek' (NWO)
and this research was supported by EC FP6 funding (contract no.
FP6-2004-IST-003673). This publication reflects the views of the
authors and not necessarily those of the EC. The Community is not
liable for any use that may be made of the information contained
herein.


\begin{thebibliography}{}

\bibitem{Roukes} M.L. Roukes, Phys. World {\bf 14}, 25 (2001).

\bibitem{Cleland} A.N. Cleland, \textit{Foundations of Nanomechanics} (Springer, Berlin, 2002).

\bibitem{Shekhter} L.Y. Gorelik \emph{et al.},
Phys. Rev. Lett. {\bf 80}, 4526 (1998).

\bibitem{Blick} A. Erbe, C. Weiss, W. Zwerger, and R.H. Blick,
Phys. Rev. Lett. {\bf 87}, 96106 (2001).

\bibitem{Flensberg} S. Braig and K. Flensberg,
Phys. Rev. B {\bf 68}, 205324 (2003).

\bibitem{Tuominen} K.D. McCarthy, N. Prokof'ev, and M.T. Tuominen,
Phys. Rev. B {\bf 67}, 245415 (2003).

\bibitem{Blanter} Ya.M. Blanter, O. Usmani, and Yu.V. Nazarov,
Phys. Rev. Lett. {\bf 93}, 136802 (2004); {\bf 94}, 049904(E) (2005).

\bibitem{Belzig} N.M. Chtchelkatchev, W. Belzig, and C. Bruder,
Phys. Rev. B {\bf 70}, 193305 (2004).

\bibitem{Mitra} A. Mitra, I. Aleiner and A.J. Milles, Phys. Rev. B {\bf 69}, 245302 (2004).

\bibitem{von Oppen} J. Koch and F. von Oppen, Phys. Rev. Lett. {\bf 94}, 206804 (2005).

\bibitem{Park C60} H. Park \emph{et al.},
Nature (london) {\bf 407}, 57 (2000).

\bibitem{C140} A.N. Pasupathy \emph{et al.},
Nano Lett. {\bf 5}, 203 (2005).

\bibitem{Kotthaus} E.M. Weig \emph{et al.},
Phys. Rev. Lett. {\bf 92}, 46804 (2004).

\bibitem{Reulet} B. Reulet \emph{et al.},
Phys. Rev. Lett. {\bf 85}, 2829 (1999).

\bibitem{Sazanova} V. Sazonova \emph{et al.},
Nature (London) {\bf 431}, 284 (2004).

\bibitem{RBM Brian} B.J. LeRoy, S.G. Lemay, J. Kong, and C. Dekker,
Nature (London) {\bf 432}, 371 (2004).

\bibitem{Nygard} J. Nyg\aa rd and D.H. Cobden, Appl. Phys. Lett. {\bf 79}, 4216 (2001).

\bibitem{Grabert} H. Grabert and M.H. Devoret, Eds. \textit{Single Charge Tunneling} (Plenum, New York, 1992).

\bibitem{Leo_Rep} L.P. Kouwenhoven, D.G. Austing, and S. Tarucha,
Rep. Prog. Phys. {\bf 64}, 701 (2001).

\bibitem{Cobden} D.H. Cobden \emph{et al.}, Phys. Rev. Lett. {\bf 81}, 681 (1998).

\bibitem{Serge} S. Lemay \emph{et al.},
Nature (London) {\bf 412}, 617 (2001).

\bibitem{DresselhausAndoEgger} M.S. Dresselhaus and P.C. Eklund,
Adv. Phys. {\bf 49}, 705 (2000). H. Suzuura and T. Ando, Phys.
Rev. B {\bf 65}, 235412 (2002). A. De Martino and R. Egger, Phys.
Rev. B {\bf 67}, 235418 (2003).

\bibitem{Note} The twisting mode vibration of the nanotube has a comparable energy with the stretching
mode. However, the twisting mode does not couple
\cite{DresselhausAndoEgger}.

\bibitem{Landau-Lifshitz} L.D. Landau and E.M. Lifshitz, \textit{Theory of Elasticity} (Pergamon, Oxford, 1986).

\bibitem{Sapmaz} S. Sapmaz, Ya.M. Blanter, L. Gurevich, and H.S.J. van der Zant,
Phys. Rev. B {\bf 67}, 235414 (2003).

\bibitem{Chemistry book} G.C. Schatz, and M.A. Ratner,
\textit{Quantum Mechanics in Chemistry} (Prentice-Hall, Englewood
Cliffs, 1993).

\bibitem{Maarten} K.C. Nowack and M. Wegewijs, cond-mat/0506552.

\bibitem{note-flensberg} K. Flensberg, private communication.

\bibitem{Abrikosov} A. A. Abrikosov, \textit{Fundamentals of the Theory of Metals} (North-Holland, New York, 1988).

\bibitem{Fedorets} D. Fedorets (unpublished).

\end{thebibliography}
\end{document}